\documentclass[]{elsarticle} 
\usepackage{graphicx}
\usepackage{dcolumn}
\usepackage{amsmath}
\usepackage{array}
\usepackage{bm}
\usepackage{textcomp}

\journal{Physics Letter B}

\begin{document}


\title{$\eta-\eta^\prime$ Mixing in the Flavor Basis and Large N}

\author[UV]{Vincent Mathieu}\corref{cor1}
\ead{vincent.mathieu@umons.ac.be}
\author[UV]{Vicente Vento}
\ead{vicente.vento@uv.es}
\cortext[cor1]{Corresponding author}
\address[UV]{Departament de F\'{\i}sica Te\`orica and Institut de F\'{\i}sica Corpuscular,\\
Universitat de Val\`encia-CSIC, E-46100 Burjassot (Valencia), Spain.}

\begin{abstract}
The mass matrix for $\eta-\eta^\prime$ is derived in the flavor basis at ${\cal O}(p^4)$ of the chiral Lagrangian using the large $N$ approximation. Under certain assumptions, the mixing angle $\phi=41.4^\circ$ and the decay constants ratio $f_K/f_\pi=1.15$ are calculated in agreement with the data. It appears that the FKS scheme arises as a special limit of the chiral Lagrangian.  Their mass matrix  is obtained without the hypothesis on the mixing pattern of the decay constants.
\end{abstract}

\begin{keyword}
meson \sep mixing
\end{keyword}


\maketitle
\section{Introduction}
The complexity of the $\eta-\eta^\prime$ mixing was investigated in many theoretical works~\cite{Leutwyler:1996sa,Gerard:2004gx,Schechter:1992iz,Feldmann:1998vh, Escribano:2008wi}. Guided by symmetry, the chiral Lagrangian is one of the most powerful tools to study the interaction of Goldstone bosons. The $\eta^\prime$ is not a Goldstone bosons because of the $U(1)$ axial anomaly but its inclusion in the chiral Lagrangian can be done using large $N$ arguments~\cite{Rosenzweig:1979ay,Witten:1980sp}.

The chiral lagrangian at next-to-leading order ${\cal O}(p^4)$ was derived  for the octet in~\cite{Gasser:1984gg} and, for the nonet, using a common expansion in ${\cal O}(1/N)={\cal O}(p^2)$ in order to reduce the number of relevant terms~\cite{Leutwyler:1996sa}. Another convention for approximating the expansion could be to keep just the leading order in $1/N$ term for any given order in momenta. Arguing that dynamics and large $N$ are independent, such an expansion was proposed~\cite{Gerard:2004gx} and leads to a good description of  $\eta-\eta^\prime$ properties in terms of a few low energy constants~\cite{Gerard:2004gx,Schechter:1992iz}.

In this work, we rewrite the mass matrix in terms of  physical quantities related to low energy constants which help us identify the relevant flavor basis well suited to deal with the mixing problem. In this basis, every unknown can be expressed (approximately) in term of four masses $M_\pi$, $M_K$, $M_\eta$, $M_{\eta^\prime}$, which become our only inputs.

The relevance of  flavor is driven by the nonorthogonal transformation between bare and physical fields~\cite{Schechter:1992iz}. This point was also identified soon after \cite{Schechter:1992iz} in the celebrated Feldmann-Kroll-Stech (FKS) formalism~\cite{Feldmann:1998vh}. To derive their mass matrix, the authors of ref.~\cite{Feldmann:1998vh} assumed that the decay constants follow the particle state mixing pattern. Under this assumption, they extracted a theoretical determination of the mixing angle.

Without the need for a  decay constants pattern, we derive from the chiral lagrangian a mass matrix in the flavor basis which compares well with FKS and show that the assumption of ref.~\cite{Feldmann:1998vh} does not influence the mass matrix and  therefore confirm, from the chiral lagrangian approach, the FKS prediction for the the mixing angle.

The presentation is as follows in the next section we review $\eta-\eta'$ mixing in the chiral Lagrangian to ${\cal O}(p^2)$ in order to show the basic ingredients of our development, namely the chiral expansion, the use of the flavor basis and its relation to the FKS formalism. In Section 3 we improve the discussion by the use of different decay constants for non-strange and strange mesons. For this purpose, we need to explore the chiral Lagrangian at ${\cal O}(p^4)$. We introduce some physically motivated assumptions which lead to accurate predictions for the mixing angle and the ratio of $f_K/f_\pi$.  In Section 4 we establish the relation between our investigation and the FKS formalism and we extract some conclusions in Section 5.

\section{The chiral Lagrangian at ${\cal O}(p^2)$} \label{sec:chiral} %
Let us briefly review what is known from chiral Lagrangians at leading order. Guided by symmetry principles, we can build an effective Lagrangian describing the low-energy behavior of QCD.
In the large $N$ limit, the relevant degrees of freedom are the nine Goldstone bosons of the symmetry breaking $U(3)_L\otimes U(3)_R\to U(3)_V$. We collect the Goldstone mesons
in a nonlinear parametrization $U=\exp\left(i\sqrt{2}\pi/f\right)$ transforming according to
\begin{equation}\label{}
    U\stackrel{G}{\longrightarrow} LUR^\dag, \; L\in U(3)_L, \; R\in U(3)_R,
\end{equation}
with $\pi = \pi^a\lambda_a$ ($\lambda_0\equiv\bm1_3\sqrt{2/3}$) or, in term of physical particles (denoted the $U(3)$ basis hereafter)
\begin{equation}\label{eq:pseudocalar}
    \pi = \sqrt{2} \begin{pmatrix}\frac{\pi^0}{\sqrt{2}}+\frac{\eta_8}{\sqrt{6}}+\frac{\eta_0}{\sqrt{3}} & \pi^+ & K^+ \\
    \pi^- & \frac{\eta_8}{\sqrt{6}}-\frac{\pi^0}{\sqrt{2}}+\frac{\eta_0}{\sqrt{3}} & K^0 \\
    K^- & \bar K^0 & -\sqrt{\frac{2}{3}}\eta_8+\frac{\eta_0}{\sqrt{3}}\end{pmatrix}.
\end{equation}

The chiral Lagrangian is constructed {\it via} an expansion in the momenta. At each order in $p^2$ we only retain retain the dominant term in the $1/N$ expansion~\cite{Gerard:2004gx}. To lowest order in derivatives, the kinetic term reads
\begin{equation}\label{eq:chiral_Lag1}
   {\cal L}^{(p^2)} = \frac{f^2}{8}\left<\partial_\mu U^\dag\partial^\mu U\right>.
\end{equation}
The other possible kinetic term
\begin{equation}\label{kinetic1/N}
    \Delta{\cal L}^{(p^2)}_{1/N} = \epsilon_1\frac{f^2}{8}\left<U^\dag\partial_\mu U\right>\left<U\partial^\mu U^\dag\right> 
\end{equation}
involves two traces and is then suppressed by one power of $1/N$~\cite{Gerard:2004gx}. The consideration of this term would lead to nondiagonal flavor transition in the currents.

Explicit symmetry breaking has to be introduced and we retain the leading order in the $1/N$ expansion at each order in $p^2$.
At leading order ($p^0$), the relevant term is the anomaly~\cite{Rosenzweig:1979ay,Witten:1980sp}
\begin{equation}\label{eq:lndet2}
 {\cal L}^{(p^0)} = \frac{\alpha_0}{2N}\left[\frac{f}{4}\left<\ln\left(\frac{\det U}{\det U^\dag}\right)\right>\right]^2 = -\frac{1}{2}\alpha_0\eta_0^2,
\end{equation}
which breaks the $U(1)_A$ symmetry and is responsible for the large $\eta'$ mass.

At order $p^2$ we must include the other symmetry breaking term provided by the quark masses. The mass term which mimics the one in the QCD Lagrangian is
\begin{equation}
    \Delta{\cal L}^{(p^2)}= \frac{f^2}{8} B\left<{\cal M} U^\dag  + U {\cal M}^\dag\right>.
\end{equation}
${\cal M}$ being  the mass matrix transforming like $U$. We use isospin $SU(2)$ symmetry and therefore ${\cal M}=\text {diag}(\tilde m,\tilde m,m_s)$.

Collecting all terms at ${\cal O}(p^2)$ and expanding the Lagrangian in the fields, one obtains
\begin{equation}\label{eq:Lag1}
    {\cal L}^{(p^2)} = \frac{1}{2}\partial_\mu\pi^a\partial^\mu\pi^b\delta_{ab} - \frac{1}{2}B\pi^a\pi^b\langle\lambda_a\lambda_b {\cal M}\rangle -\frac{1}{2}\alpha_0\eta_0^2,
\end{equation}
from which one can read the physical  masses (squared) $m^2_\pi =B\tilde m$ and $m^2_K =B(\tilde m + m_s)/2$.

$f$ is the pion decay constant $f_\pi= 132$ MeV as it can be deduced from the conserved current $A_\mu^a = -f\partial_\mu \pi^a$ and the definition
\begin{equation}\label{}
    \left<0|A_\mu^a(x)|\pi^b\right> = -i f_\pi p_\mu \delta^{ab} e^{-ipx}.
\end{equation}
In Lagrangian \eqref{eq:chiral_Lag1}, all Goldstone bosons have the same decay constant $f=f_\pi$.

The incorporation of the isosinglet $\eta_0$ induces a mixing with the $\eta_8$. The masses of the two physical states $\eta$ and $\eta^\prime$ are the eigenvalues of the mass matrix for the $\eta_8-\eta_0$ system
\begin{equation}\label{eq:massmatrix80}
    {\cal M}_{80}^2 = \frac{1}{3}\begin{pmatrix}4m_K^2-m_\pi^2 & -2\sqrt{2}(m_K^2-m_\pi^2) \\
    -2\sqrt{2}(m_K^2-m_\pi^2) & 2m_K^2+m_\pi^2 + 3\alpha_0\end{pmatrix}.
\end{equation}
In the $SU(3)_F$ limit, where all quarks have the same mass, {\it i.e.} $m_s=\tilde m$, the coupling between the singlet and the octet disappears. At this stage, $\alpha_0$ is an unknown parameter.

Furthermore, as we shall see,  the diagonal mass matrix induces a privileged basis. For this it is useful to work in the flavor basis. Moreover,  the chiral lagrangian  results compare nicely with the  FKS formalism~\cite{Feldmann:1998vh} in this basis. For this purpose, we use the representation
\begin{equation}\label{eq:Uflavor_basis}
    \pi = \sqrt{2}\text{ diag }\left(u\bar u,d\bar d,s\bar s\right).
\end{equation}
We dropped the noninteresting nondiagonal fields.

In the flavor basis, the mass matrix reads~\cite{Gerard:2004gx}
\begin{equation}\label{}
    {\cal M}^2_{uds} = \frac{\alpha_0}{N} \begin{pmatrix} 1&1&1\\1&1&1\\1&1&1+R\\\end{pmatrix}
     + m^2_\pi \begin{pmatrix} 1&0&0\\0&1&0\\0&0&1\\\end{pmatrix},
\end{equation}
with $R = (2N/\alpha_0)(m^2_K-m^2_\pi)$. The mixing is provided only by the anomaly as expected from the ideal mixing between the vector particles $\omega$ and $\phi$.

We next introduce the fields $\eta_s = s\bar s$ and $\eta_q = (u\bar u+d\bar d)/\sqrt{2}$ (and the orthogonal counterpart of $\eta_q$, $(u\bar u-d\bar d)/\sqrt{2}$ which decouples under isospin symmetry) allowing to rewrite the mass matrix as
\begin{equation}\label{}
    {\cal M}^2_{qs} = \begin{pmatrix} m^2_\pi + 2(\alpha_0/N) &(\alpha_0/N)\sqrt{2}\\(\alpha_0/N)\sqrt{2}& 2m^2_K -m^2_\pi+ (\alpha_0/N)\\\end{pmatrix}.
\end{equation}
We easily obtain a condition on the two eigenvalues $E_1^2$ and $E_2^2$ of this matrix~\cite{Gerard:2004gx,Georgi:1993jn}
\begin{equation}\label{}
    \frac{E_1^2-m^2_\pi}{E_2^2-m^2_\pi} = \frac{3+R-\sqrt{9-2R+R^2}}{3+R+\sqrt{9-2R+R^2}} \leq \frac{3-\sqrt{3}}{3+\sqrt{3}} = 0.268.
\end{equation}
Hence, as shown by Georgi~\cite{Georgi:1993jn}, this mixing scheme cannot provide the physical ratio $(m^2_{\eta'}-m^2_\pi)/(m^2_\eta-m^2_\pi)= 0.313$. We have only one parameter to reproduce two physicial states, this is clearly not enough.

We investigate in the next Section an improvement consisting in the use of different decays constants for nonstrange and strange mesons. For this purpose, we need to explore the chiral Lagrangian at ${\cal O}(p^4)$.  Another mechanism for improvement, the inclusion a third mainly gluonic state, was investigated in~\cite{Mathieu:2009sg}.

\section{The chiral Lagrangian at ${\cal O}(p^4)$} \label{sec:chiral}
For our purposes, we are only interested in  terms contributing to the kinetic part, the mass matrix and the decay constants. We restrict ourselves to terms involving only zero or two derivatives, and we only keep the leading order in $1/N$. There are three terms of interest at ${\cal O}(p^4)$ and they  involve a single trace over flavor \cite{Gerard:2004gx}
\begin{equation}\label{eq:lag_Dp4}
\begin{split}
\Delta{\cal L}^{(p^4)}=& \frac{f^2}{8}\bigg[-\frac{B}{\Lambda^2}\left< {\cal M}\partial_\mu\partial^\mu U^\dag\right> + \frac{B^2}{2\Lambda_1^2}\left<{\cal M}U^\dag{\cal M}U^\dag\right> \\ &+\frac{B}{2\Lambda_2^2}\left<{\cal M}U^\dag\partial_\mu U\partial^\mu U^\dag\right>\bigg]
+\text{h.c.}
\end{split}
\end{equation}

The three low energy constants enter the observables without a clear physical meaning. $\Lambda$ and $\Lambda_2$ induce a splitting between the $\pi$ and $K$ decay constants. $\Lambda_1$ and $\Lambda_2$ enter in the corrections to the mass matrix. Fitting their values on observables leads to a consistent $\eta-\eta'$ scheme at ${\cal O}(p^4)$ \cite{Gerard:2004gx}.

We are interested in a more physical interpretation of the low energy constants and we  aim at an analytical resolution of the mass matrix. We therefore would like to reduce the number of parameters since a two-by-two matrix gives us only two independent equations.

The basic hypothesis to reduce the numbers of parameters in the FKS scheme was to neglect the transition between different quark flavors. In other words, a condition that reads
\begin{equation}\label{eq:noflavor_trasition}
    J_\mu^q\left|\eta_s\right>=J_\mu^s\left|\eta_q\right>=0
\end{equation}
This condition is automatically satisfied in the large $N$ approximation since Eq. (\ref{kinetic1/N}) is suppressed.

We are now interested in expressing the mass matrix in term of more physical quantities like the decay constants. To this aim, we rotate away the $\Lambda$ term by a chiral transformation at ${\cal O}(p^2)$~\cite{Gerard:2004gx}
\begin{equation}\label{}
    U\longrightarrow U^\prime=U-\frac{B}{2\Lambda^2}({\cal M}-U{\cal M}^\dag U).
\end{equation}
Such a redefinition preserves the unitarity of $U$ up to ${\cal O}(p^4)$ and amounts the simple substitutions
\begin{subequations}
\begin{eqnarray}
  \frac{1}{\Lambda_1^2} &\to& \frac{1}{\Lambda_1^2} + \frac{1}{\Lambda^2},\\
  \frac{1}{\Lambda_2^2} &\to& \frac{1}{\Lambda_2^2} + \frac{2}{\Lambda^2},
\end{eqnarray}
\end{subequations}
in the Lagrangian \eqref{eq:lag_p4}, but does not preserve the anomalous term~\eqref{eq:lndet2}. The Lagrangian after rotation reads
\begin{equation}\label{eq:lag_p4}
\begin{split}
{\cal L}^{(p^4)} = {\cal L}^{(p^2)} &+  \frac{f^2}{8}\left[\frac{B^2}{2}\left(\frac{1}{\Lambda_1^2}+\frac{1}{\Lambda^2}\right)\left<{\cal M}U^\dag{\cal M}U^\dag\right> \right.  \\  &  \left.+\frac{B}{2}\left(\frac{1}{\Lambda_2^2}  +\frac{2}{\Lambda^2}\right)\left<{\cal M}U^\dag\partial_\mu U\partial^\mu U^\dag\right>\right] +\text{h.c.}  \\
 &  -\frac{\alpha_0}{2N}\frac{B}{\Lambda^2}\frac{f^2}{8}\left<{\cal M}^\dag U - U^\dag{\cal M}\right>\left<\ln\left(\frac{\det U}{\det U^\dag}\right)\right>.
\end{split}
\end{equation}
Where ${\cal L}^{(p^2)}$, given in~\eqref{eq:Lag1}, collects the three terms (kinetic energy and two explicit symmetry breaking terms) at ${\cal O}(p^2)$.

The physics  described by the Lagrangian~\eqref{eq:lag_p4} is exactly the same as that of the original Lagrangian. But now, the low-energy constants appear in a more convenient way. Indeed, to derive the mass matrix, one should always bring the kinetic terms into its canonical form by mean of a redefinition of the fields. With our particular Lagrangian~\eqref{eq:lag_p4}, the wave-function renormalizations are simply proportional to the matrix of the decay constants. This matrix is proportional to $\left\langle \lambda_a\lambda_b{\cal M}\right\rangle$. Renormalizing the fields is easy in the case of a diagonal matrix since it only amounts to renomalizing the fields by a simple rescaling without any rotation. This is not the case in the $U(3)$ basis where the rotation is mandatory \cite{HerreraSiklody:1997kd} to brings us in the flavor basis where $\left\langle \lambda_a\lambda_b{\cal M}\right\rangle$ becomes diagonal. It is then more advantageous to start directly in the flavor basis~\eqref{eq:Uflavor_basis}. In the flavor basis the kinetic terms
\begin{equation}\label{eq:kinetic_F}
    \frac{1}{2}\left(\frac{f_q}{f}\right)^2\partial_\mu\eta_q\partial^\mu\eta_q + \frac{1}{2}\left(\frac{f_s}{f}\right)^2\partial_\mu\eta_s\partial^\mu\eta_s,
\end{equation}
does not present any mixing term. Moreover, since the kinetic energy and the decay constants follow the same pattern, the matrix of the decay constants is also diagonal. This avoids unwanted (and sometimes overlooked \cite{Klopot:2009cm}) transition elements $J^8_\mu|\eta_0\rangle$. The particular structure of the mass matrix imposes then a privileged basis as claimed below and used in the FKS scheme \cite{Feldmann:1998vh}.

The flavor decay constants can be easily expressed in term of the physical ones (in the $U(3)$ basis):
\begin{subequations}\label{decay_constants}
\begin{eqnarray}
  f^2_q &=& f^2_\pi, \\
  f^2_s &=& 2f_K^2-f^2_\pi.
\end{eqnarray}
\end{subequations}
The mass matrix  then involves those physical decay constants. Introducing the parameter $y =f_q/f_s$, we can derive from the Lagrangian~\eqref{eq:lag_p4} the mass matrix in a more convenient form
\begin{equation}\label{eq:MqsNLO1}
    {\cal M}_{qs}^2 = \begin{pmatrix} M^2_{qq} + 2\alpha &\alpha y \sqrt{2}\\ \alpha y \sqrt{2}& M^2_{ss}+ \alpha y^2\\\end{pmatrix} + {\cal O}\left(\frac{\alpha}{\Lambda^2}\right).
\end{equation}
We have defined for convenience $\alpha=(\alpha_0/N)(f/f_q)^2$.

${\cal O}\left(\alpha/\Lambda^2\right)$ stands from the contribution of the last term in~\eqref{eq:lag_p4}, i.e.
\begin{equation}\label{eq:new_term}
      \left<{\cal M}^\dag U - U^\dag{\cal M}\right>\left<\ln\left(\frac{\det U}{\det U^\dag}\right)\right> \propto (\tilde m\sqrt{2}\eta_q + m_s\eta_s )(\sqrt{2}\eta_q + \eta_s).
\end{equation}
Assuming a negligible term ${\cal O}\left(\alpha/\Lambda^2\right)$  provides us analytical formulas, while the inclusion of this term leads to a unavoidable numerical procedure.
Comparing our mass matrix with one derived in~\cite{Gerard:2004gx}, we see that the parameter $\tilde\delta$ arising in their expressions, disappeared for a more physical parameter $y$.

From \eqref{eq:MqsNLO1} and neglecting the term ${\cal O}\left(\alpha/\Lambda^2\right)$ (or equivalently taking the limit $\Lambda\to\infty$), it is straightforward to extract the value of the parameters $y$ and $\alpha$ in function of the masses. Equating the determinant and the trace of \eqref{eq:MqsNLO1} with the mass matrix matrix of the physical states, diag($M^2_\eta,M^2_{\eta'}$), we obtain
\begin{eqnarray}
    y^2 &=& 2\frac{M^2_\eta M^2_{\eta^\prime}-M^2_{ss}(M^2_\eta +M^2_{\eta^\prime} - M^2_{ss})}{M^2_\pi(M^2_\eta +M^2_{\eta^\prime}-M^2_\pi)-M^2_\eta M^2_{\eta^\prime}}, \\
    \alpha &=& \frac{M^2_\eta +M^2_{\eta^\prime} -M_\pi^2 - M^2_{ss}}{2+y^2},\\
    \sin2\varphi &=& \frac{2\sqrt2\alpha y}{M^2_{\eta^\prime} - M^2_\eta}.
\end{eqnarray}
We quoted the expression for the mixing angle in the flavor basis defined in a standard way by
\begin{equation}\label{eq:mixing}
    \begin{pmatrix} \eta\\ \eta' \end{pmatrix} =
    \begin{pmatrix}\cos\phi & -\sin\phi\\ \sin\phi & \cos\phi\end{pmatrix}
    \begin{pmatrix} \eta_q\\ \eta_s \end{pmatrix}.
\end{equation}

In the mass matrix \eqref{eq:MqsNLO1}, $M_{qq}$ and $M_{ss}$ are the unknown masses of the pseudoscalar $q\bar q$ and $s\bar s$ states. They can be related to physical masses by expressing them in the $U(3)$ basis in analogy with~\eqref{decay_constants}. We obtain~\cite{Gerard:2004gx}
\begin{subequations}\label{masses}
\begin{eqnarray}
  M^2_{qq} &=& M^2_\pi, \\
  M^2_{ss} &=& 2M_K^2-M^2_\pi + (M_K^2-M^2_\pi)^2\left(\frac{2}{\Lambda_1^2} - \frac{1}{\Lambda_2^2}\right).
\end{eqnarray}
\end{subequations}
We are not able to relate $M^2_{ss}$ with the physical masses at ${\cal O}(p^4)$. We can approximate $M^2_{ss}$ by $2M_K^2 - M_\pi^2$ (exact only at ${\cal O}(p^2)$), the error on $M_{ss}$ being less than $5\%$. This approximation  coincides with the value used in the FKS paper~\cite{Feldmann:1998vh}.

We then obtain the value for the two parameters $y$ and $\alpha$ and a prediction for the physical quantities
\begin{equation}\label{eq:predictions}
 \frac{f_K}{f_\pi} =  1.146, \qquad
  \phi = 41.40^\circ.
\end{equation}
Our calculated values~\eqref{eq:predictions} lie in the usual range $[40^\circ,45^\circ]$ \cite{pheno_value} and are in agreement with the data on $\eta$ and $\eta^\prime$ decays (other value for the mixing angle can be predicted theoretically~\cite{other_value}). The phenomenological value for the ratio $f_K/f_\pi=1.193(0.009)$~\cite{PDG} is also close to our predicted value.

The approximation $M^2_{ss}=2M_K^2 - M_\pi^2$ is only valid at leading order and receives a correction at order ${\cal O}(p^4)$~\cite{Gerard:2004gx}. The extra contribution reduces the value of the mixing angle and $f_K/f_\pi$ as shown on Fig.~1 and 2.

\begin{figure}[htb]\begin{center}
  \includegraphics[width=0.5\linewidth]{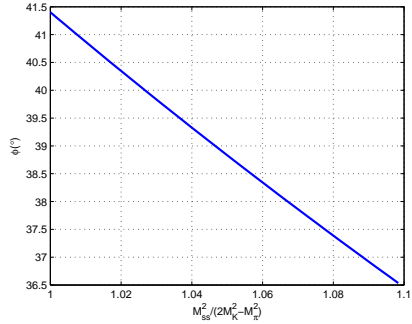}
    \caption{Mixing angle in the flavor basis as a function of $M^2_{ss}/(2M_K^2 - M_\pi^2)$.}\label{fig:phi}
\end{center}
\end{figure}

\begin{figure}[htb]\begin{center}
  \includegraphics[width=0.5\linewidth]{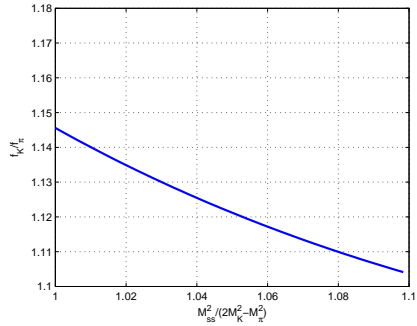}
    \caption{$f_K/f_\pi$ as a function of $M^2_{ss}/(2M_K^2 - M_\pi^2)$.}\label{fig:y}
\end{center}
\end{figure}

\section{Comparison with FKS Formalism}
\label{sec:FKS}
We next compare the prediction of the chiral Lagrangian with the FKS results~\cite{Feldmann:1998vh}. The basic hypothesis of the FKS formalism was the assumption that decay constants in the flavor basis follow the same mixing pattern~\eqref{eq:mixing} of the states
\begin{equation}\label{FKS_hyp}
    \begin{pmatrix} f_\eta^q & f_\eta^s \\ f_{\eta'}^q & f_{\eta'}^s\end{pmatrix} =
    \begin{pmatrix}\cos\phi & -\sin\phi\\ \sin\phi & \cos\phi\end{pmatrix}
    \begin{pmatrix} f_q & 0 \\ 0 & f_s\end{pmatrix}.
\end{equation}
Under this assumption, the mass matrix is derived from current algebra and reads
\begin{equation*}
    {\cal M}^2_{\text{FKS}} = \begin{pmatrix}
    M^2_{qq}+\frac{\sqrt{2}}{f_q}\langle0|\frac{\alpha_s}{4\pi}G\tilde G|\eta_q\rangle & \frac{1}{f_s}\langle0|\frac{\alpha_s}{4\pi}G\tilde G|\eta_q\rangle\\
    \frac{\sqrt{2}}{f_q}\langle0|\frac{\alpha_s}{4\pi}G\tilde G|\eta_s\rangle & M^2_{ss}+\frac{1}{f_s}\langle0|\frac{\alpha_s}{4\pi}G\tilde G|\eta_s\rangle. \end{pmatrix}
\end{equation*}
Let us see how it arises from the chiral Lagrangian.

In our effective approach, the anomaly, \emph{via} $\Delta{\cal L}^{(p^0)}$ in \eqref{eq:lndet2}, couples equally to each flavor {\it before} the renormalization of the fields. Indeed we have
\begin{equation}\label{}
    \left(\partial^\mu A_\mu^0\right)_{\text{anomaly}} = f\alpha_0(\sqrt{2}\eta_q+\eta_s),
\end{equation}
which should be compare to $\left(\partial^\mu A_\mu^0\right)_{\text{anomaly}} = 3\alpha_s/(4\pi)G_{\mu\nu}^a\tilde G^{\mu\nu}_a$.
After a proper normalization of the fields, $\langle0|\eta_q|\eta_q\rangle=f/f_q$, see~\eqref{eq:kinetic_F}, we find
\begin{eqnarray}
  \langle0|\frac{\alpha_s}{4\pi}G\tilde G|\eta_q\rangle &=& \alpha\sqrt{2}f_q\\
  \langle0|\frac{\alpha_s}{4\pi}G\tilde G|\eta_s\rangle &=& \alpha y^2f_s
\end{eqnarray}
Rewriting ${\cal M}^2_{\text{FKS}}$ in our notation, we get (in their paper~\cite{Feldmann:1998vh}, the authors denoted the anomaly by $a^2=\alpha$)
\begin{equation}\label{eq:FKS}
    {\cal M}^2_{\text{FKS}} = \begin{pmatrix}
    M^2_{qq}+2\alpha & \alpha y \sqrt{2} \\
    \alpha y \sqrt{2} & M^2_{ss}+\alpha y^2\end{pmatrix}
\end{equation}
We clearly see the equivalence of the mass matrices \eqref{eq:FKS} and \eqref{eq:MqsNLO1} at ${\cal O}(\alpha/\Lambda^2)$. The mass matrix is then symmetric and we do not need to impose the equality of the off diagonal terms by hand as in ref.~\cite{Feldmann:1998vh}.

The important point is that we did not use any assumptions of the mixing scheme for the decay constants in the derivation of \eqref{eq:MqsNLO1}. The hypothesis \eqref{FKS_hyp} is not mandatory from our effective field theory point of view and is not physically justified on general grounds. By comparing both approaches, we notice that the hypothesis on the decay constants pattern is not consistent with the chiral Lagrangian at next to leading order in the large $N$ limit since it amounts to neglecting the first term in \eqref{eq:lag_Dp4}.

\section{Conclusion}
Keeping only the leading term in $1/N$ at each order in the momentum expansion, we derived at ${\cal O}(p^4)$ a mass matrix in the flavor basis. The quantities are expressed in term of the masses and the ratio of decay constants $y$, instead of the low energy constants, allowing a clear understanding. Such a form allows the extraction of two parameters $\phi$ and $y$ from mass inputs. Their values are in agreement with the data.

The choice of the flavor basis is motivated the simple expression of the kinetic term and the diagonal matrix for the decay constants. Such features, not apparent in the $U(3)$ basis, make the calculation tedious in that basis and render the physics more obscure.

The properties of the flavor basis were previously identified and analyzed in ref.~\cite{Schechter:1992iz} and in the FKS scheme~\cite{Feldmann:1998vh}. In this work, we relate this formalism with the chiral Lagrangian approach where no assumption on the mixing scheme for the decay constants is required. It is shown that the FKS Ansazt is not compatible with the chiral Lagrangian. The difference stands in the parameter $\Lambda$. Taking the limit $\Lambda\to\infty$ renders the FKS scheme whereas its value, extracted from the data, is of the same magnitude as those of the other low-energy constants~\cite{Gerard:2004gx}.

The predicted values \eqref{eq:predictions} are in agreement with the data but leave a room for improvement since assumptions were invoked. Moreover, the glue content in the $\eta'$ wave function should be properly considered at ${\cal O}(p^4)$. To this aim, the inclusion of the pseudoscalar glueball in the chiral Lagrangian in under construction \cite{Mathieu:2009sg,futur}

\section*{Acknowledgements}
We thank J.-M. G\'erard for instructive discussions. V.M. thanks the hospitality of the UMons Physics Department of C. Semay and the CPAN for financial support.
This work was supported in part  by HadronPhysics2,  a FP7-Integrating Activities and Infrastructure Program of the
European Commission under Grant 227431, by the MICINN (Spain) grant FPA2007-65748-C02- and by GVPrometeo2009/129.

\end{document}